\documentclass[11pt,a4paper]{article}
\usepackage{authblk}
\usepackage{graphicx}
\usepackage{float}

\usepackage[margin=1in]{geometry}
\usepackage{abstract}

 \newcommand{\EF}{$E_\mathrm{F}$}
\newcommand{\EB}{$E_\mathrm{B}$}

\newcommand{\kF}{$k_\mathrm{F}$ }
\newcommand{\kz}{$k_\mathrm{z}$}
\newcommand{\hn}{$\mathrm{h\nu}$}

 \newcommand{\GM}{$\overline{\Gamma\mathrm{M}}$}
 \newcommand{\GK}{$\overline{\Gamma\mathrm{K}}$}
 \newcommand{\MGM}{$\overline{\mathrm{M}\Gamma\mathrm{M}}$\ }
 \newcommand{\KGK}{$\overline{\mathrm{K}\Gamma\mathrm{K}}$\ }
\newcommand{\mgm}{$\overline{\mathrm{M}\Gamma\mathrm{M}}$}

 \newcommand{\ZGZ}{$\mathrm{Z}\Gamma\mathrm{Z}$ }

\newcommand{\aGeTe}{$\alpha$\nh GeTe(111)\ }

 \newcommand{\n}[1]{$n$\nobreakdash-\hspace{0pt}}
\newcommand\nh{\mbox{-}}
\newcommand\degC{$\,^{\circ}\mathrm{C}$}
\newcommand\dg{$^{\circ}$}
\newcommand{\kspace}{\textbf{k}-space}
\newcommand{\kA}{$\mathrm{\AA}^{-1}$}

\newcommand{\Pz}{$P_{z}$}
\newcommand{\Pxy}{$\overrightarrow{P}_{x,y}$}
\newcommand{\Px}{$P_{x}$}
\newcommand{\Py}{$P_{y}$}
\newcommand{\lll}{$\langle 111 \rangle$}

\title{Surface versus bulk contributions to the giant Rashba splitting in the ferroelectric \aGeTe semiconductor.}

\author[1]{J. Krempask{\'y}\thanks{juraj.krempasky@psi.ch}}
\author[2,3]{ H.~Volfov{\'a}}
\author[1,6]{S.~Muff}
\author[1]{N.~Pilet}
\author[1,5]{G.~Landolt}
\author[1,8]{M.~Radovi{\'c}}
\author[1]{M.~Shi}
\author[4]{D.~Kriegner}
\author[4]{V.~Hol{\'y}}
\author[3]{J.~Braun}
\author[3]{H.~Ebert}
\author[1]{F.~Bisti}
\author[1]{V.A.~Rogalev}
\author[1]{V.N.~Strocov}
\author[7]{G.~Springholz\thanks{Gunther.Springholz@jku.at}}
\author[2,3]{J.~Min{\'a}r\thanks{jan.minar@cup.uni-muenchen.de}}
\author[1,6]{J.~H.~Dil\thanks{hugo.dil@epfl.ch}}

\affil[1]{Swiss Light Source, Paul Scherrer Institut, CH-5232 Villigen PSI, Switzerland}
\affil[2]{New Technologies-Research Center University of West Bohemia, Plze{\v n}, Czech Republic}
\affil[3]{Department of Chemistry, Ludwig Maximillian University, 81377 Munich, Germany}
\affil[4]{Department of Condensed Matter Physics, Charles University in Prague, Ke Karlovu 5, 121 16 Praha 2, Czech Republic}
\affil[5]{Physik-Institut, Universit{\"a}t Z{\"u}rich, Winterthurerstrasse 190, 8057 Z{\"u}rich, Switzerland}
\affil[6]{Institute of condensed matter physics, Ecole Polytechnique F{\'e}d{\'e}rale de Lausanne, CH-1015 Lausanne, Switzerland}
\affil[7]{Institut f{\"u}r Halbleiter-und Festk{\"o}rperphysik, Johannes Kepler Universit{\"a}t, A-4040 Linz, Austria}
\affil[8]{SwissFEL, Paul Scherrer Institut, CH-5232 Villigen PSI, Switzerland}

\begin{document}
\date{}

\maketitle

\begin{abstract}

\noindent
\textbf{In systems with broken inversion symmetry spin-orbit coupling (SOC) yields a Rashba-type spin splitting of electronic states, manifested in a k\nh dependent splitting of the bands. While most research had previously focused on  2D electron systems~\cite{Hugo_TR, Okuda:2013}, recently a three-dimensional (3D) form of such Rashba-effect was found in a series of bismuth tellurohalides BiTe\textit{X} (\textit{X}=I, Br, or Cl)~\cite{BiTeI_NC,BiTeI_Hugo,BiTeCl_Gabriel,BiTeI_Crepaldi}. Whereas these materials exhibit a very large spin-splitting they lack an important property concerning functionalisation, namely the possibility to switch or tune the spin texture.  This limitation can be overcome in a new class of functional materials displaying Rashba-splitting coupled to ferroelectricity: the ferroelectric Rashba semiconductors (FERS)~\cite{Picozzi_AdvM,Picozzi_Front}. Using spin- and angle-resolved photoemission spectroscopy (SARPES) we show that \aGeTe forms a prime member of this class, displaying a complex spin\nh texture for the Rashba\nh split surface and bulk bands arising from the intrinsic inversion symmetry breaking caused by the ferroelectric polarization of the bulk (FE). Apart from pure surface and bulk states we find \hbox{surface-bulk} resonant states (SBR) whose wavefunctions entangle the spinors from the bulk and surface contributions. At the Fermi level their hybridization results in unconventional spin topologies with cochiral helicities and concomitant gap opening. The \aGeTe surface states (SS) and SBR states make the semiconductor surface conducting. At the same time our SARPES data confirm that GeTe is a narrow-gap semiconductor with a well resolved Dirac point in the 3D-valence band and a giant spin\nh splitting $\Delta k_R \approx 0.11$ \AA$^{-1}$ of the bulk states. These results show that the \aGeTe surface/bulk electronic states are endowed with spin properties that are theoretically challenging to anticipate. As the helicity of the spins in Rashba bands is connected to the direction of the FE polarization, this work paves the way to all\nh electric non\nh volatile control of spin-transport properties in semiconductors.}

\end{abstract}

\noindent\textrm{IV-VI} semiconductor compounds such as lead chalcogenides, SnTe or GeTe have been extensively studied by band-structure calculations for which relativistic effects and SOC were included~\cite{calcs1,Cardona_PE_book,Cardona_book}. Emphasis was primarily put on the band-edge properties and the fundamental gap~\cite{Cardona_PE_book} rather than the spin properties. However, recent theoretical work has predicted that in topologically trivial \aGeTe the bulk bands are expected to be fully spin-polarized with the spin chirality directly coupled to the FE polarization~\cite{Picozzi_AdvM,Picozzi_Front}. A further development of this concept into applications requires first of all an experimental verification and, deeper theoretical considerations.
\begin{figure}[h!]
\centering
\includegraphics[width=0.9\textwidth]{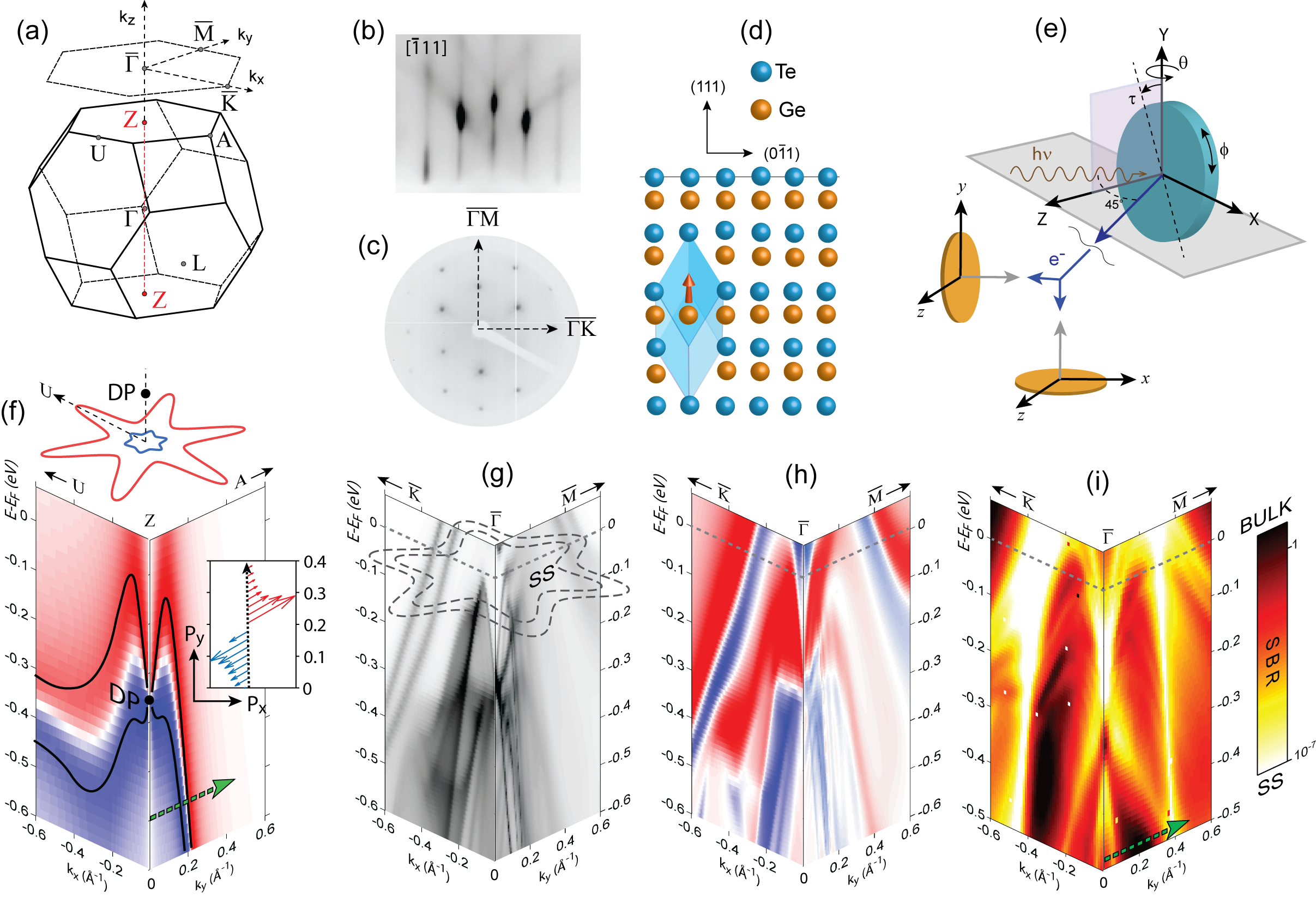}
    \caption{\textbf{Basic properties of \aGeTe and experimental geometry.} \textbf{(a)} Brillouine zone; \textbf{(b)} RHEED and \textbf{(c)} LEED pattern of the bulk and surface, respectively;
\textbf{(d)} Te-terminated semi-infinite surface structural model with highlighted distorted rhombohedral cell. Orange arrow shows the displacement of the Ge-atom responsible for the polar FE ordering. \textbf{(e)} Experimental geometry with schematic drawing of our SARPES experiment. \textbf{(f)} Spin-resolved initial state Bloch spectral functions calculated for bulk Brillouine zone along ZU and ZA directions. (inset) $\protect\overrightarrow{P}_{x,y}$ quiver plot along the green arrow. Calculated spin-integrated \textbf{(g)} and spin-resolved \textbf{(h)} BSF from a semi-infinite surface. \textbf{(i)} Logarithmic false color map of 1SM determinant demonstrating bulk, surface and SBR character of calculated ARPES spectral intensities~\cite{JBraun_1SM,Echenique_Pendry}. The green arrow indicates the $k_y$ momentum span where bulk Rashba splitting is best observed.}
\label{F1}
\end{figure}
\newline
\indent
Experiments were performed on 200 nm thick \aGeTe monocrystalline films grown by molecular beam epitaxy on BaF$_2$(111) substrates~\cite{BaF2_1,BaF2_2}, with a Te-terminated surface prepared \textit{in situ} [see Methods and Supplemental Information (SI) \textrm{I,II}.1]. Fig.~1(b-d) show the surface structural model together with results of the sample characterization by \textit{in\nh situ} low-energy electron diffraction (LEED) and reflection high-energy electron diffraction (RHEED). The lack of surface reconstruction in LEED and the presence of Kikuchi lines in RHEED indicate superior quality \aGeTe surface with spontaneous FE moments~\cite{Springholz_PRL}. Below T$_\mathrm{C}$ = 700 K there is a spontaneous paraelectric to FE phase transition in which the cubic rock\nh salt phase changes to a polar rhombohedral structure, forming a FE\nh phase induced by elongation of the unit cell along the \lll~direction (see the orange arrow in Fig.~1d). A concomitant shift in the Te-sublattice is inducing an almost perfect collective FE\nh order, up to $\approx$92\% oriented along the \lll~direction perpendicular to the sample surface  (SI \textrm{II}.2). The resulting symmetry breaking inside the \aGeTe triple-layers is responsible for the bulk Rashba-splitting. Of importance for the surface\nh sensitive UV\nh SARPES on unpoled samples is that the spontaneous polarization of FE domains persist in the sub\nh surface region, as this ensures a net spin polarization also for the bulk states (SI \textrm{II}.3).
\newline
\indent
Due to the small probing depth of ARPES, measured energy-, angle- and spin-resolved spectra can hardly be compared directly with a calculated band\nh structure because photoemission intensities do not relate directly to dispersions of bare bands.  The task to describe the surface electronic structure differs from that in the bulk in several ways. Due to the broken crystal symmetry the bulk spin-resolved band-structure calculations is different compared to a semi-infinite surface [Fig.~1(f-g)], mainly due to the presence of surface states (SS). The star-like feature in a constant energy map (CEM) from bulk initial states orients along \GK, whereas surface-like initial states orient along \GM. Furthermore, for certain binding energies and momenta, the canted spin of the bulk states is opposite to what is predicted for surface states with rather tangential helicity. Thus, mixing the spectral weights from surface and bulk states inevitably result in a competition of spin textures between bulk and surface states which in our case facilitate their discrimination in a spin-resolved experiment. 
\newline
\indent
Since ARPES data critically depend on photoemission final states, to build a bridge between SARPES and theoretical band\nh structure calculations we use so-called one\nh step photoemission model (1SM, see Methods). In this most rigorous interpretation besides bulk states also surface evanescent (SS) and surface-bulk-resonance (SBR) states are considered. To better understand their spectral weights, we employ the 1SM determinant criterion~\cite{Echenique_Pendry,JBraun_1SM} in Fig.~1(i) in order to qualitatively discern their contributions in the measured valence states. Within the determinant logarithmic scale small values correspond to surface states with the maximum of their wavefunctions between the first atomic layer and the surface potential. States which embed in the bulk continuum have values close to one and are shown by black shades. On the other hand the SBR states, being partially reflected from bulk states, are given as red shades. The SS/SBR and SBR/bulk transitions appear gradually, but as a whole the SBR is the primary spectral weight dispersing across the gap, forming metallic states at \EF, and degenerating with bulk Rashba bands at lower binding energies (\EB). 
\newline 
\indent 
\begin{figure}[h!]
\centering
\includegraphics[width=1\textwidth]{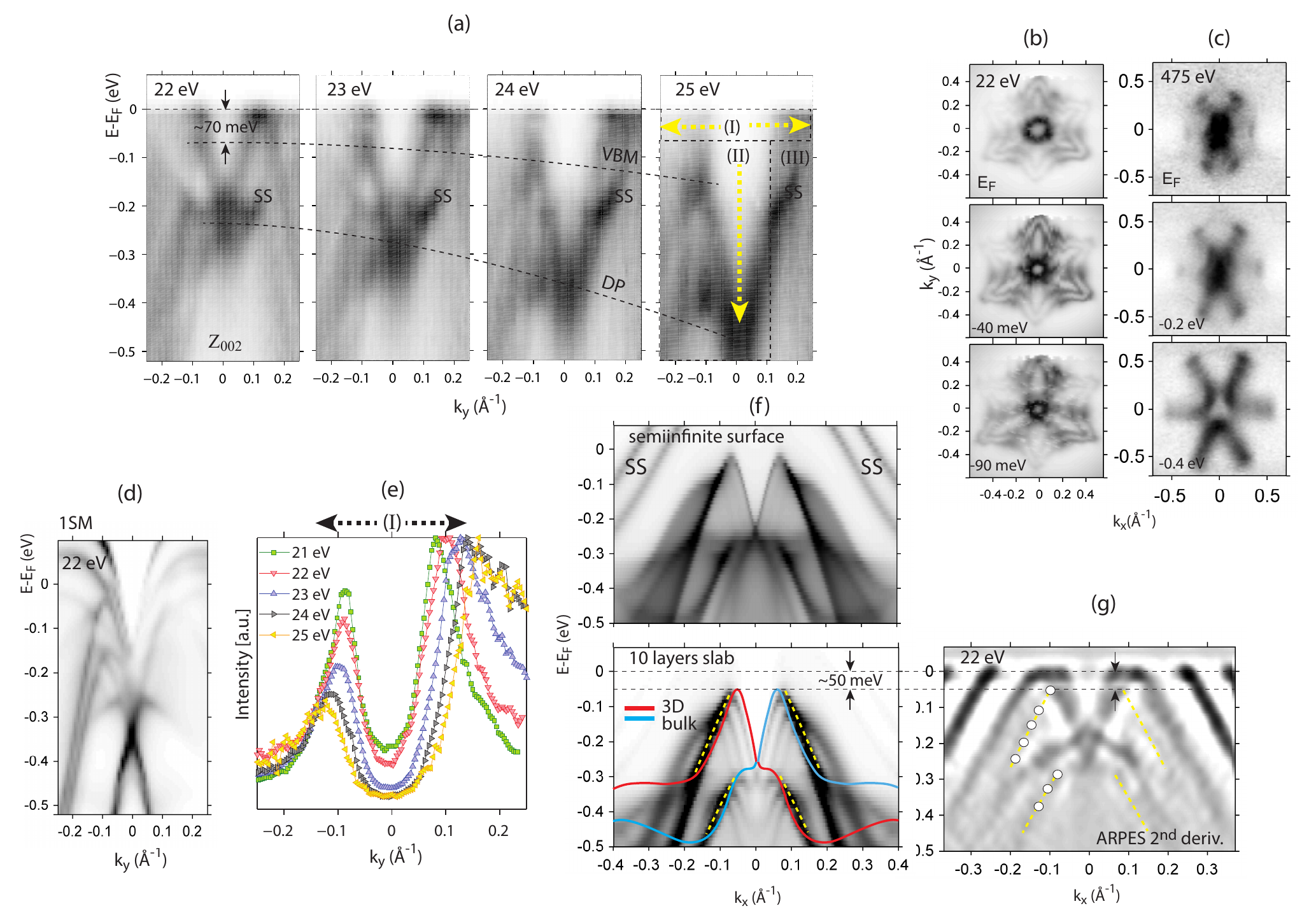}
\caption{\textbf{ARPES \kz~dispersion and comparison with semi-infinite surface calculations.} (\textbf{a}) 
UV\nh ARPES data taken along the \ZGZ direction for selected photon energies point out the Z-point around 22 eV with the Dirac point (DP) located around -0.25 eV \EB. The dashed lines  highlight the dispersion of the DP along with the top of the valence band, indicating a bulk gap of few tens meV. Yellow arrows indicate the 2D-like dispersion of the SBR inside the gap (section \textrm{I}), in contrast to the 3D-like dispersion of the bulk-like band in section \textrm{II}. (\textbf{b}) CEMs for selected \EB~inside the gap measured at \hn~of 22 eV and 475 eV in (\textbf{c}). \textbf{(d)} ARPES 1SM calculation calculated for \hn~of 22 eV along \MGM showing the influence of matrix\nh element effects on the spectral weight, which expose the surface states in the band-map section \textrm{III}. \textbf{(e)} MDCs showing the 2D-like dispersion of SBR states at \EF.
\textbf{(f)} Comparison of the semi-infinite and 10-layer slab calculations and corresponding second-derivative ARPES data in panel \textbf{(g)} highlights the SBR buildup in the valence band [yellow dashed lines in \textbf{(f)} and white dots in  \textbf{(g)}].}
\label{F2}
\end{figure}
We first concentrate on the localization of the Z-point in \kspace. The relevant region of \kz\nh momenta manifesting the Rashba\nh splitting is expected along the \ZGZ direction, denoted in red in Fig.~1(a). The corresponding band maps shown in Fig.~2(a) have pronounced spectral intensity of the surface states at \EF~and the Dirac point (DP), which we used to relate experimental data to band-structure calculations. The DP has minimum of -0.25 eV \EB~around a \hn~of 22 eV, constituting the Z-point along the \ZGZ direction. The VBM at the Z-point is $\approx$-70 meV \EB~in \MGM and $\approx$-50 meV \EB~in \KGK mirror plane, as seen in Fig.~2(a,f). 
\newline
\indent
What is the origin of the metallic states at \EF? We note that SBR materialize as a superposition of bulk and surface spectral weight contributions, readily identified in corresponding theoretical CEMs in Fig.~1(f-g). We note that by mixing SS with bulk states the SS typically lose their well-defined surface character at \EF~by forming a SBR. Our results suggest that hybridization between the SS and bulk states is in-fact "mediated" by SBR. Recently such hybridization mechanism was also confirmed in Sb$_2$Te$_3$ topological insulators~\cite{Seibel_PRL_Sb2Te3}. An important observation is that the SBR fill the gap from the valence band maximum (VBM) all the way up to \EF, with progressively increased bulk-like spectral weight toward the VBM [Fig.~2(b)]. The in-gap states reaching \EF~display rather 2D than 3D character in \kz, as seen in the band-map section \textrm{I} in Fig.~2(a). The metallic states filling the gap couple more to surface than bulk states for two reasons. First, consistently with the 1SM predictions in Fig.~1(i), the spectral weight at \EF~for momenta \kF$<$0.3 \kA~belongs to the SBR states and not to pure surface states. This corroborates their momentum distribution dispersing toward the pure surface states near \hbox{\kF$\approx$0.4 \kA}, as seen by the momentum distribution curves (MDC) in Fig.~2(e) and Fig.~3b. Second, the SBR states filling the gap clearly detach from the main 3D-valence band dispersive trend inside the band-map section \textrm{II}.  
\begin{figure}[h!]
\centering
\includegraphics[width=1\textwidth]{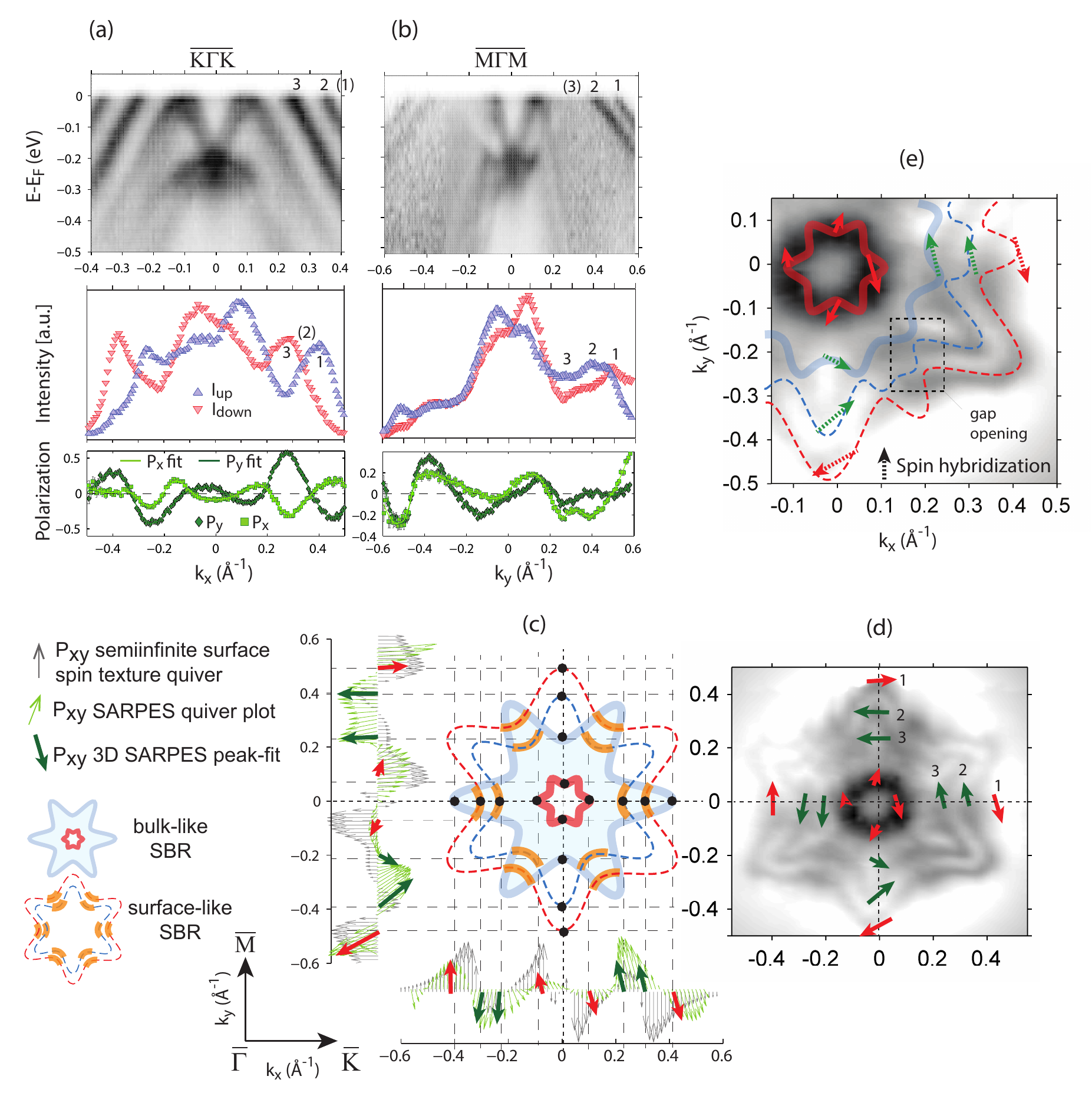}
\caption{\textbf{Energy dispersions measured by ARPES and SARPES near \EF.} (\textbf{a,b}) (top) Valence band ARPES data with populations of spin-up and spin-down electrons near \EF, as a function of electron momentum along \KGK and \mgm (mid panels)[ with corresponding spin polarizations \Px~and \Py~at \EF~(bottom panels). (\textbf{c}) Expected spin-resolved Fermi surface map (FSM) compared with experimental data in \textbf{(d)}. Spin-resolved theoretical predictions from semi-infinite surface in Fig.~1(h) (gray arrows), are overlaid with smoothed $\protect\overrightarrow{P}_{x,y}$~data from \textbf{a,b} (light-green arrows). Dark-green arrows mark the results from the 3D vectorial spin analysis based on simultaneously fitted MDC peaks [thin lines in \textbf{(d)}, see SI \textrm{V}]. The difference in spin-texture between the simplified model sketched in panel \textbf{(c)} and the experimental result is due to spin hybridization of the SBR wavefunctions indicated by dashed arrows in panel \textbf{(e)} (see text).}
\label{F3}
\end{figure}
\newline
\indent
To further understand our ARPES data it is useful to compare them with the semi-infinite and finite slab of 10-layers. As seen in Fig.~2(f), it is evident that the SBR-states pile up as a superposition of bulk-projected states giving rise to enhanced spectral intensity. Such SBR spectral intensity buildup is highlighted with dashed lines and is dominant in the valence band ARPES. It refers to SBR states with strong coupling to the bulk continuum. Consequently, in agreement with the 1SM calculations, whole portions of pure bulk\nh VB states, denoted with red and blue lines, are overlaid in ARPES with the SBR states making their identification difficult by means of ground state bulk band-structure calculations~\cite{Picozzi_AdvM,Cardona_PE_book}. On the other hand the comparison between the 1SM ARPES calculation for \hbox{\hn=22 eV} in Fig.~2(d) with the corresponding experimental data in panel (a) is rather convincing. 
\newline
\indent
We discuss next our spin-resolved measurements near \EF. Spin-resolved populations of spin-up and spin-down electrons were measured along the $\tau$ direction perpendicular to the scattering plane [Fig.~1(e)]. We use a well-established quantitative 3D-vectorial analysis~\cite{Fabian_NJP} for simultaneous peak fitting of spin-integrated MDCs and spin polarizations. 
The polarization curves are modeled until the best fit is reached by simultaneously fitting the MDC intensity and the polarizations \Px, \Py~and \Pz~(SI \textrm{III}). Of importance are the MDC-peaks 1-2-3 labeled in Fig.~3(a,b), the visibility of which is comprehensively described within the simultaneous fitting. The resulting spin-texture at \EF~is seen in Fig.~3(c). Thick arrows compare the 3D-fits with the in\nh plane polarization predictions (\Pxy) from the semi-infinite surface (quiver plot in gray color). As a whole the SARPES data at \EF~suggest that the SBR metallic states filling the gap have unconventional spin texture with parallel spin orientations different from the simple Rashba model. Our conjecture is that such spin topology in the momentum distribution can be reconciled via SOC-induced hybridization of the SBR-states. Highlighted in orange are regions where the bulk and surface contributions to the SBR wavefunctions hybridize. The resulting gap opening in momentum distribution, sketched in Fig.~3(e), is inducing a small spectral weight shift which realize the cochiral spin texture (SI \textrm{III}). Similar interband gap opening and spin reorientation has been observed in 2D Rashba systems~\cite{ISOC_PRL,Bartosz_NJP}. As a matter of fact, the in-gap states inside the band-map section \textrm{I} in Fig.~2(d) display rather 2D than 3D character.
\newline
\indent
To further examine our ARPES data we note that our experimental geometry predicts in 1SM calculations a suppression of spectral weight in \MGM below the DP for positive momenta. In fact, on the experimental side the bulk contributions to the SBR function wash out; uncovering the surface states inside the non-dispersive band-map section \textrm{III} [Fig.2~(a)]. This photoemission\nh transition matrix\nh element effect provides a direct evidence that the bulk and surface states indeed overlap in the initial states. As seen in Figure 2, a side effect of the vanished spectral weight intensity is that the CEMs appear with pronounced 3-fold rather than 6-fold symmetry~\cite{note3-6,Cardona_book}. 
\begin{figure}[h!]
\includegraphics[width=1\textwidth]{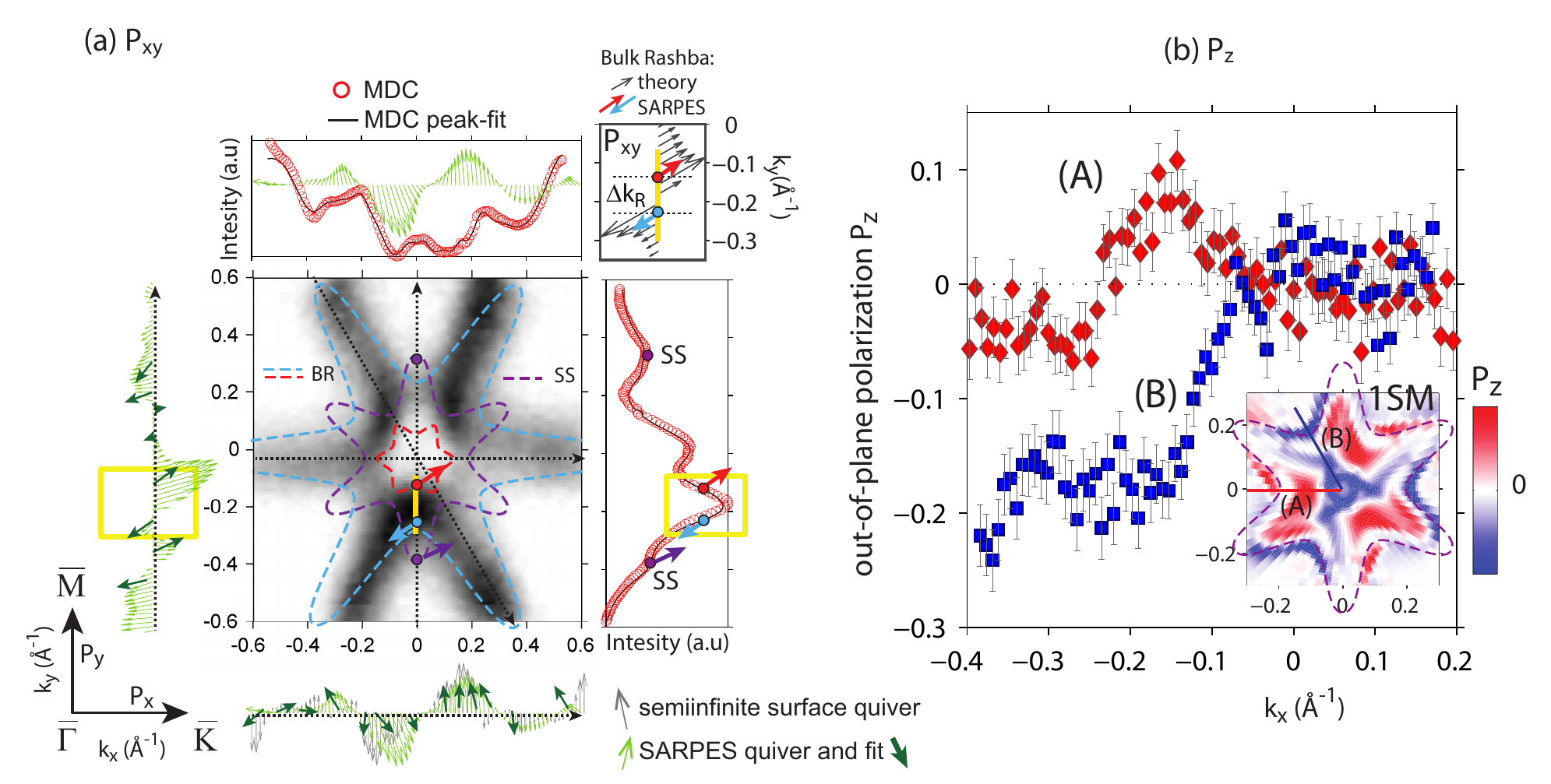}
\caption{\textbf{Rashba splitting and out-of-plane spin polarization of bulk bands.} \textbf{(a)} ARPES constant energy map (CEM) measured in the Z-point ($h\nu=$475 eV) at -0.4 eV \EB.  Spin-resolved MDCs are measured along \MGM (vertical) and \KGK (horizontal) measured in equivalent Z-point at $h\nu=$22 eV. The spectral weight of bulk Rashba states for positive momenta vanishes as seen in Fig.~3(a), exposing the surface states (SS) denoted with violet arrows. The SS helicity (violet arrows) is opposite to pure Rashba bulk states (arrows with red and blue hues). Yellow frames highlight momenta where the 1SM predicts pure bulk states (yellow arrows in Fig.~1(f,i). The measured in\nh plane $\protect\overrightarrow{P}_{x,y}$~polarization for both bulk bands aligns with canted spins consistently with the bulk Rashba initial states, showing a giant spin\nh splitting \hbox{$\Delta k_{R}\approx$0.11 \kA}. Along \KGK the semi-infinite surface quiver is displayed in gray arrows. It is compared to measured (light-green quiver) and 3D-fitted (dark-green) spin textures. 
\textbf{(b)} Hexagonal warping in the out-of-plane spin polarization \Pz~measured along the paths (A) and (B). (inset) Corresponding 1SM \Pz~calculation.} 
\label{F4}
\end{figure}
\newline 
\indent 
We now look at the influence of the opposite helicity of SS and SBR states on the measured spin texture.
It turns out that an ideal locus of \kspace~momenta for measuring pure bulk\nh Rashba properties is around -0.5 eV \EB~in \MGM. The black shades in the photoemission determinant along the green arrow in Fig.~1(i) predict bulk states with distinct canted helicity for in\nh plane \Pxy~polarization [\Px$\approx$2\Py, as seen in the inset in Fig.~1(f)]. The same canted helicity is perfectly reflected in the measured polarization in Fig.~4(a). Such experimental confirmation traces back the magnitude and direction of the spin polarization vector to the states from which these electrons originate. This is non-trivial, since especially for systems with large SOC, the total angular momentum is relevant for photoemission process rather than only the spin quantum number~\cite{Heinzmann:2012}. Since the degree of spin polarization can depend strongly on the \hn~\cite{Heinzmann:2012}, the same states were probed with different photon energies (SI \textrm{III}). Because the same canted spin arrangement was confirmed in all cases, we conclude that the measured spin signals trace back to initial state bulk Rashba spinors, making this the first experimental confirmation of the spin texture of a FERS.
\newline
\indent
Finally we comment on the out-of-plane polarization. The band anisotropy between \GM~and \GK~induce rather strong hexagonal warping effect responsible for the out-of-plane \Pz~spin polarization component~\cite{Picozzi_AdvM,Fu:2009w,Eremeev:2012}. SARPES data in data seen in Fig.~4(b) confirm the hexagonal warping measured in two equivalent \MGM directions denoted (A) and (B). As seen in the inset 1SM calculations, the measured warping is in excellent agreement with measured data.
\newline
\indent
In a broader perspective, the unveiled \aGeTe bulk spin structure has far reaching consequences both for fundamental physics and applications. Based on our results \aGeTe is the prime candidate in spin-orbit-driven Rashba physics in 3D \kspace. Resolving the spin structure at \EF~and inside the narrow gap is relevant in terms of potential device-oriented applications. Namely, GeTe appears to have always $p$-type metallic transport properties due to Ge-vacancies which lead naturally to a large density of holes in the valence band states~\cite{Edwards_PRB}. Nevertheless it manifests macroscopic, robust and reversible ferroelectricity promising for technological applications. Our findings confirm that a bulk Rashba effect and ferroelectricity coexist and are coupled. We emphasize that the giant spin\nh splitting $\Delta k_R \approx 0.11$ \AA$^{-1}$ is formed by bands which interconnect the Rashba-split conduction- and valence band continuum. We anticipate that the \aGeTe Rashba physics is likely to extend to Ge$_{1-x}$Mn$_x$Te multiferroic Rashba semiconductors because the rhombohedral distortion persists up to $x=0.25$~\cite{Springholz_PRL}. Thus the quest for semiconductors with combined functional properties might extend to systems where magnetism, ferroelectricity and Rashba physics would offer unique property combinations for potential device applications. 

\begingroup
    \fontsize{10pt}{12pt}\selectfont
\section*{Methods}
\subsection*{Sample preparation method.}
Experiments were performed on 200 nm thick \aGeTe films grown by molecular beam epitaxy on BaF$_2$(111) substrates~\cite{BaF2_1,BaF2_2}. A protective amorphous Te-cap used to avoid surface oxidation and degradation was  removed in the ultrahigh vacuum chamber by annealing the thin film for 30 min at 250\degC. The surface quality checked by LEED shows the 1$\times$1 diffraction pattern of the \aGeTe surface seen in Fig.~1c. 

\subsection*{ARPES and spin-resolved ARPES}
The VUV-ARPES/SARPES experiments were performed at the COPHEE end-station of the SIS beamline ~\cite{Hoesch_JESRP}, Swiss Light Source (SLS), using $p$-polarized photons in the energy range 20--25 eV, and an Omicron EA 125 hemispherical energy analyzer equipped with two orthogonally mounted classic Mott detectors. The whole set-up allows the simultaneous measurements of all three spatial components of the spin-polarization vector for
each point of the band structure. The geometry of the experimental set-up is schematically shown in Fig.~1e. The spin-resolved MDCs were measured by rotating the sample azimuth $\phi$ such that \MGM or \KGK were aligned perpendicular to the scattering plane. Direct and immediate visualization of the spin splitting as a function of electron momentum was measured by exploiting the manipulator tilt angular degrees of freedom [see $\tau$ in Fig.~1(e)]. The spin\nh integrated Fermi surfaces were measured with the channeltron detectors of the SARPES set-up by \hbox{on-the-fly} sweeping the manipulator tilt angle ($\tau$) for selected polar angles $\theta$ as seen in the experimental setup in Fig.~1(e). The angular and resolution-broadened energy range in spin-resolved mode were 1.5\dg and 60 meV, respectively. In spin-integrated mode the resolutions were set to 0.5\dg and 20 meV. To enhance the bulk sensitivity compared to VUV-ARPES, experiments were also performed in the soft-X-ray energy range at the ADRESS beamline~\cite{Strocov_ADRESS}. The experimental results in the main text were reproduced on three samples with individual annealing preparation over different experimental runs. All data were taken at 20 K.

\subsection*{First principle calculations}
The ab-initio calculations are based on density functional theory (DFT) 
as implemented within the multiple scattering theory (SPRKKR)~\cite{SPR13,Minar_RPP}.  SOC has been naturally included by use of a fully relativistic  four-component scheme. As s first step of our investigations we 
performed self-consistent calculations (SCF) for 3D bulk as well as 2D  semi-infinite surface of  \aGeTe within the screened KKR formalism~\cite{Minar_RPP}. The corresponding ground state band structures are presented in terms of Bloch spectral functions (BSF). The converged SCF potentials  served as an input quantities for our spectroscopic investigations. The  ARPES calculations were performed in the framework of the fully  relativistic one-step model of photoemission~\cite{JBraun_1SM,Ebert_book} in its  spin-density matrix formulation~\cite{JBraun_Rashba}, which accounts properly for the complete spin-polarization vector, in particular for Rashba systems 
like GeTe. Together with a realistic model for the surface barrier  potential, one-step calculations based on a semi-infinite half-space  configuration were decisive to substantiate the \aGeTe  surface-generated spectral features on both qualitative and quantitative  levels.

\subsection*{Acknowledgments}Constructive discussions with S. Picozzi, R. Calarco and R. Bertacco are gratefully acknowledged. This work was supported by the Swiss National Science Foundation Project No. PP00P2\_144742\/1. We acknowledge the financial support from German funding agencies DFG (SPP1666) and the German ministry BMBF (05K13WMA) is also gratefully acknowledged (H.V.,.H.E.,J.B. and J.M.).  J.M. acknowledges the CENTEM project, Reg.No. CZ.1.05/2.1.00/03.0088, co-funded by the ERDF as part of the Ministry of Education, Youth and Sports OP RDI program. D.K. and V.H. acknowledge the financial support of Czech Science Foundation (project 14-08124S)
\endgroup




\begin{thebibliography}{10}
\bibitem{note3-6}
\bibinfo{note}{A 3-fold rather than 6-fold symmetry in CEMs is expected due to
  rhombohedral distortion along the \lll~axis which for GeTe give rise to a
  structure with 3-fold rather than 6-fold coordination.}
\end{thebibliography}
\end{document}